# Mobility induced unsaturated high linear magnetoresistance in transition-metal monopnictides Weyl semimetals

**Chandra Shekhar\*, Vicky Süß and Marcus Schmidt**

Max Planck Institute for Chemical Physics of Solids, 01187 Dresden, Germany

\*Email: shekhar@cpfs.mpg.de, cgsbond@gmail.com

## Abstract

Discovery of Weyl fermions in semimetallic tansition-metal monopnictides is a major breakthrough in condensed matter physics. A Weyl semimetal is characterized by the existence of robust Weyl points and unclosed topological surface states in the form of Fermi arc. All the four compounds of the Weyl semimetal transition monopnictide family i.e. NbP, TaP, NbAs and TaAs exhibit extremely high mobility and unsaturated high magnetoresistance (MR). For example, MR values are $8.5 \times 10^5$ % at 1.85 K for NbP and $1.5 \times 10^5$ % at 3 K in 9 T for TaAs.  NbP also achieves very low value of residual resistivity 0.63 $\mu\Omega$ cm at 2 K due to suppression of scattering resulting in ultra-high mobility $5 \times 10^6$ cm$^2$/Vs, Interestingly, we find that the mobility of these compounds play an important role for such a large MR.

# 1. Introduction

Semimetals have always been very intriguing for researchers because the valence band maximum and conduction band minimum lie near the Fermi surface. This means that effect of even a small perturbation can easily be seen on the physical properties [1-6]. Recently, topological semimetals have attracted tremendous attention as it has expanded the family of topological materials in condensed matter physics and material science [7-17]. The interest in this class of materials is both academic and as a huge potential in applications. Like topological insulators, semimetals are also topologically protected nontrivial metallic phases of matter, wherein the nontrivial topological nature guarantees the existence of exotic Fermi surface as Fermi arc [9, 11, 14-16]. Due to the semimetallic character of materials, the conduction and valence bands usually cross each other near the Fermi energy and depending on the degeneracy in momentum space, the topological semimetals are further distinguished and classified into Dirac and Weyl semimetals (WSM). If the materials break the time-reversal and/or inversion symmetries, the non-degenerate band-crossing points are called Weyl points, which act as magnetic monopoles in momentum space and always come in pairs and those materials are known as Weyl semimetals. Around the Weyl points bands are linearly dispersed in three dimension. Whereas, when the time-reversal and/or inversion symmetries are protected, a pair of Weyl points are degenerate, forming another topological phase called a Dirac semimetal [18, 19]. Topological surface states of WSM are characterized by Fermi arcs which are responsible for showing exotic transport properties like chiral anomaly [20], large spin Hall conductivity [8, 21] *etc*. Search for different topological semimetallic materials has been stimulated by the recent WSM discovery in transition-metal monopnictides [10-12, 14-17]. Many other topological semimetals have been predicted such as half Heusler GdPtBi [22, 23], $Bi_{0.97}Sb_{0.03}$ [24], $ZrTe_5$ [25], and $Na_3Bi$ [26], pyrochlore iridate $Y_2Ir_2O_7$ [9], $HgCr_2Se_4$ [8], $Hg_{1-x-y}Cd_xMn_yTe$ [27], $LaBi_{1-x}Sb_xTe_3$ [28], and $TlBiS_{1-x}Se_{2-x}$ [29]. Moreover, GdPtBi, $Bi_{0.97}Sb_{0.03}$, $ZrTe_5$ and $Na_3Bi$ have been realized experimentally only from the chiral anomaly (negative magnetoresistance) and they await the direct visualization of Fermi arc.

Transition-metal monopnictides NbP, TaP, NbAs and TaAs have been known for a long time constituting a family of nonmagnetic semimetallic materials [30]. All the compounds crystalize in body-centered-tetragonal structure with nonsymmorphic space group $I4_1md$ (No. 109), which lack inversion symmetry. Figure 1(a) shows the atomic arrangements in a unit cell of this series of compounds which can be described by the arrangement of regular trigonal prisms of transition metals with pnictide atom at the center. Besides similarity in crystal structure, all four mentioned compounds share very identical electronic band structures resulting similar properties. Figures 1(b, c) illustrate the evolution of Weyl points in these compounds. In the absence of spin − orbit coupling (SOC), bands cross each other near the Fermi energy to give rise to in nodal ring protected by the mirror symmetry. In the presence of the SOC the nodal ring is gapped and now the valence and conductions bands only touch at two distinct points away from the mirror plane (figure 1(c)). The two Weyl points thus generated are protected by time reversal symmetry. Transition-metal monopnictides have 12 pairs of Weyl points in Brillouin zone. Resulting topological surface states characterized as a Fermi arc connect two Weyl points and the presence of Fermi arc is now well investigated and confirmed by the angle-resolved photoemission spectroscopy [11, 14-16] and the scanning tunneling microscopy [31-33] in this family. Here, we report a first proven family of Weyl semimetals that exhibit unsaturated linear MR and ultra-high mobility. These linear MR follow classic model of Paris and Littlewood wherein mobility governs MR confirming that high MR is expected in high mobility compounds [34].

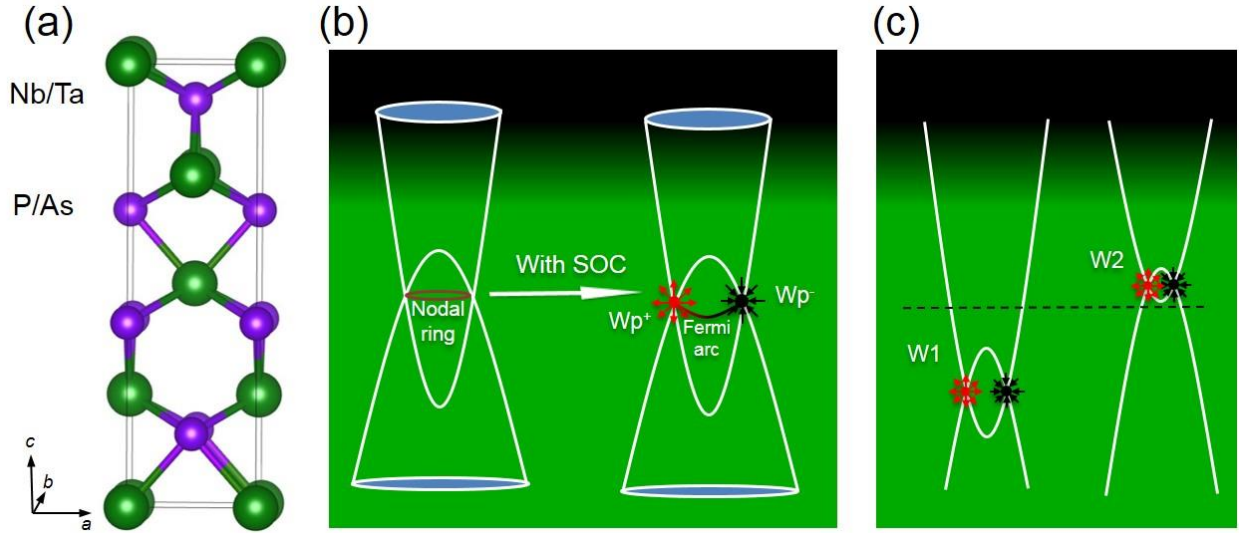

**Figure 1.** (a) A unit cell of transition metal–monopnictides, which breaks the center of inversion symmetry. (b) Band schematic of a semimetal forming nodal ring without spin–orbit coupling (SOC) at intersection region of valance and conduction bands (left panel). The birth of Weyl semimetal by SOC (right panel). $Wp^+$ and $Wp^-$ are Weyl points, which are connected by topological surface states forming Fermi arc. (c) In transition metal– monopnictides, a pair of Weyl points located in electron pocket (W1) as well as in hole pocket (W2) and black line is charge neutral line (Fermi energy).

## 2. Experimental details

Single crystals of NbP, TaP, NbAs and TaAs were grown via chemical vapour transport [35, 36]. As a first step for polycrystalline material, stoichiometric quantities of Nb, Ta (Alfa Aesar, 99.99%), P (Alfa Aesar, 99.999%) and As (Chempur, 99.9999%) were weighed in accurately in a quartz ampoule, flushed with Ar, sealed under vacuum and heated the sealed quart tube in two consecutive temperatures of 600 °C for 24 h and 800 °C for 24 h. In the next step for crystals growth, we used microcrystalline powders from step one and then added iodine (7 − 8 mg/ ml) before sealing the powders in quartz tube. The crystal growth was carried out in a two-zone furnace between 900 – 1050 °C for 2-4 weeks. Here, $I_2$ act as a transport agent. To obtain high quality of the crystals, temperature gradient is one of the most important parameters and it varies on material to material. We optimized the temperature gradient for each material mentioned in present investigation and it is 900 °C (source) – 1000 °C (sink) for NbP, TaP and 900 °C (source) – 1050 °C (sink) for NbAs, TaAs. More details about crystal growth and their characterization can be found in our recent studies [12, 13, 17, 37]. The experimental lattice parameters of the four compounds are $a = 3.3314(8)$ Å, $c = 11.3649(3)$ Å for NbP, $a = 3.3166(9)$ Å, $c = 11.3304(6)$ Å for TaP, $a = 3.4492(6)$ Å, $c = 11.6647(3)$ Å for NbAs and $a = 3.4310(4)$ Å, $c = 11.6252(6)$ Å for TaAs, in close agreement with previous reports [30, 38]. Resistivity measurements were performed in PPMS using the AC and DC mode of the AC-Transport options. Our well characterized crystals were cut into bar-shapes using wire saw keeping long direction parallel to the crystallographic $a$-axis. The physical dimensions (length × width × thickness) are $3.1 \times 1.6 \times 0.56$ mm$^3$ for NbP, $1.1 \times 0.42 \times 0.16$ mm$^3$ for TaP, $0.93 \times 0.83 \times 0.17$ mm$^3$ for NbAs and $1.5 \times 0.42 \times 0.28$ mm$^3$ for TaAs. Resistivity and Hall resistivity measured were done in four and five-probe geometries respectively in constant current source of $3 – 4$ mA.

## 3. Results and discussion

Resistivity, $\rho_{xx}$ of all crystals at zero field decreases with decreasing temperature as expected from metallic compound. Their low residual resistivity and high residual resistivity ratio values reflect high quality of crystals [12, 13]. Besides well-established Weyl property, these compounds are also well known for producing ultrahigh mobility and ultrahigh magnetoresistance (MR). We now focus on the MR measurements. In presence of the magnetic fields (B), the nonzero transverse currents will experience a Lorentz force in the inverse-longitudinal direction. Such a back flow of carriers eventually increases the apparent longitudinal resistance, resulting in the extremely high MR. The MR is commonly calculated as the ratio of the change in resistivity due to the applied magnetic field and the initial resistivity from the relation $MR = \rho_{xx}(B) - \rho_{xx}(0)/\rho_{xx}(0)$ while mobility is estimated by taking the slope of Hall resistivity at high magnetic field region, where it is almost linear with field at different temperatures. The measured MR of NbP, TaP, NbAs and TaAs at selected temperatures are shown in figure 2 and its remarkable features are: (i) Positive and unsaturating MR that shows systematic variations with temperature and field. (ii) MR is almost constant up to 50 K while it decreases sharply above 50 K. (iii) The more striking feature in MR is that all compounds show Shubnikov de-Haas (SdH) oscillations below 20 K, reflecting the high quality of the crystals. The detailed analysis of SdH oscillations and their fermiology can be found elsewhere [12, 13, 39, 40]. Among the four compounds, NbP exhibits highest MR of $8.5 \times 10^5$ % at 1.85K in a field of 9 T and interestingly does not show any sign of saturation under ultrahigh fields of 62 T at 1.5 K [12]. This value is five times larger than the value that reported for the same field in WTe$_2$ [41], another WSM. The MR of NbP is as high as 250% even at room temperature and 9 T (inset of figure 2a). Moreover, the other members of this series also exhibit similar order of MR like NbP at all temperature and field ranges. From above results, we can say that these materials have unique property of high and never saturating MR.

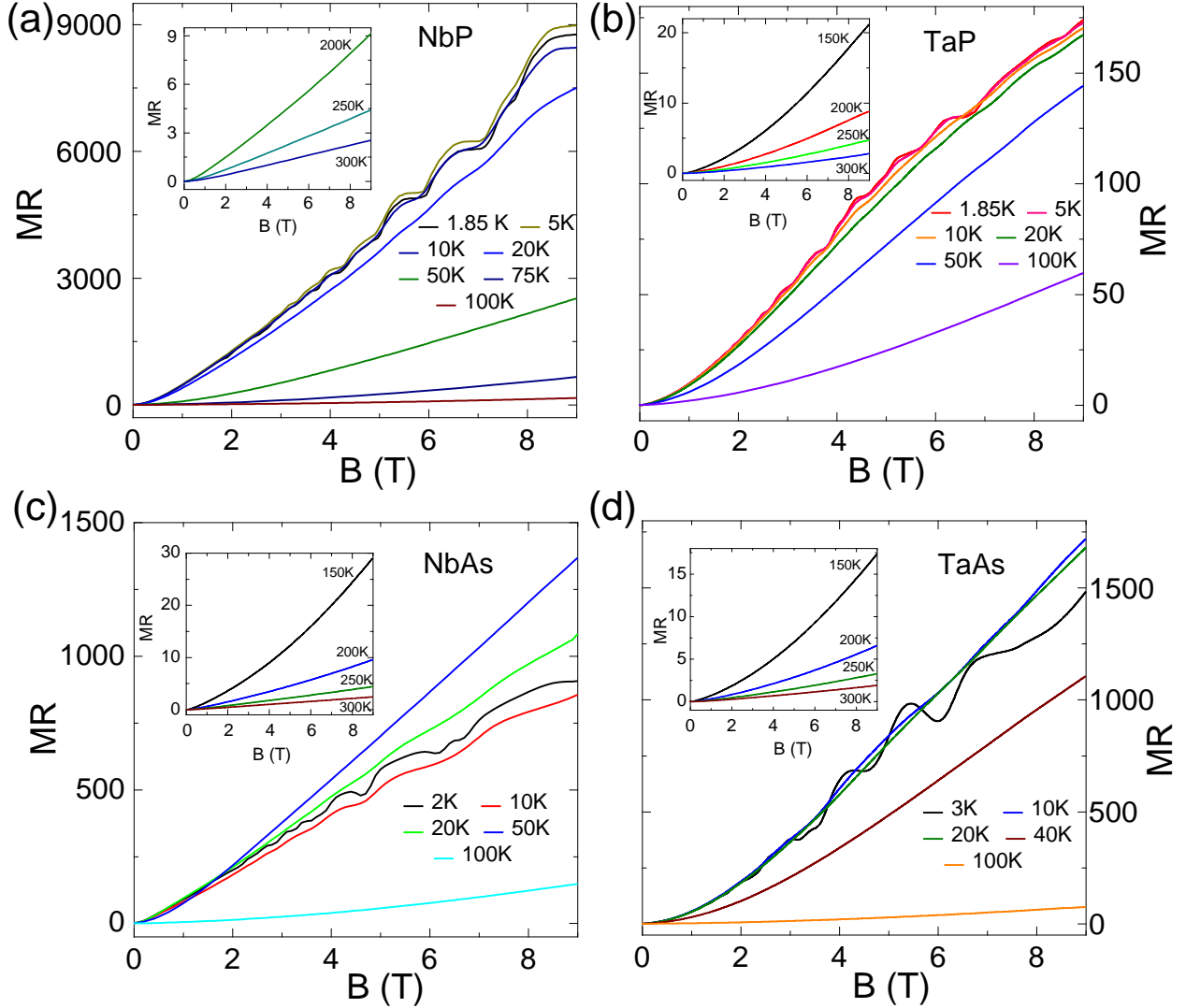

**Figure 2.** Calculated transverse magnetoresistance at different temperatures from the field dependent resistivity using $MR = \rho_{xx}(B) - \rho_{xx}(0) / \rho_{xx}(0)$ for (a) NbP, (b) TaP, (c) NbAs, (d) TaAs. Inset show their respective magnetoresistance at high temperature $\geq 150$ K.

We now turn our discussion to the Hall resistivity, $\rho_{xy}$. The Hall resistivity is defined $\rho_{xy} = V_y \cdot t / I_x$ , where $V_y$ is the voltage developed normal to both magnetic field and current and $t$ is the thickness of the crystal. The carrier mobility and concentration usually derived from $\rho_{xy}$ are two important parameters of a material. We have performed Hall measurements in positive and negative field directions to improve the accuracy of data. For the sake of simplicity, we used the single-carrier Drude model, $\mu_{avg}(T) = R_H(T)/\rho_{xx}(T)$ , where $\mu_{avg}(T)$ is the average mobility and $R_H(T)$ is the Hall coefficient calculated from the linear slope of the $\rho_{xy}(T)$ at high field. However, there is a change in the sign of Hall coefficient from negative to positive between 125 − 170 K depending on the material e.g. this temperature is about 125 K for Nb-compounds and 170 K for Ta-compounds [12, 13]. The charge carrier density lies between $10^{18} - 10^{20}$ cm$^{-3}$ in the temperature range 2 − 300 K. The observed small carrier density at low temperature and its huge change on increasing the temperature are a typical semimetallic nature. A large MR is usually associated with high mobility and it plays a major role in the charge transport in a material and consequently determines the efficiency of devices. We calculated the average mobility and plotted it against temperature in figure 3. NbP shows the highest mobility $5 \times 10^6$ cm$^2$V$^{-1}$s$^{-1}$ and the lowest residual resistivity 0.63 μΩ cm at 2 K. These values are very close to that of Cd$_3$As$_2$ [42]. Mobilities of others members are $3 \times 10^5$ cm$^2$V$^{-1}$s$^{-1}$ for TaP, $5 \times 10^5$ cm$^2$V$^{-1}$s$^{-1}$ for NbAs and $4 \times 10^5$ cm$^2$V$^{-1}$s$^{-1}$ for TaAs at 2 K and their residual resistivities are 3.2 μΩ cm for TaP, 6.2 μΩ cm for NbAs and 4.2 μΩ cm for TaAs. From the above values of mobility and residual resistivity, it is clear that the anomalously low residual resistivity can be achieved in a clean system where the mobility attains ultrahigh value. When the magnetic field turns on, low temperature resistivity steeply increases, resulting in the extremely large MR [12, 13]. We find that the average mobility and change in MR correlate very well for the entire temperature range for all the four compounds.

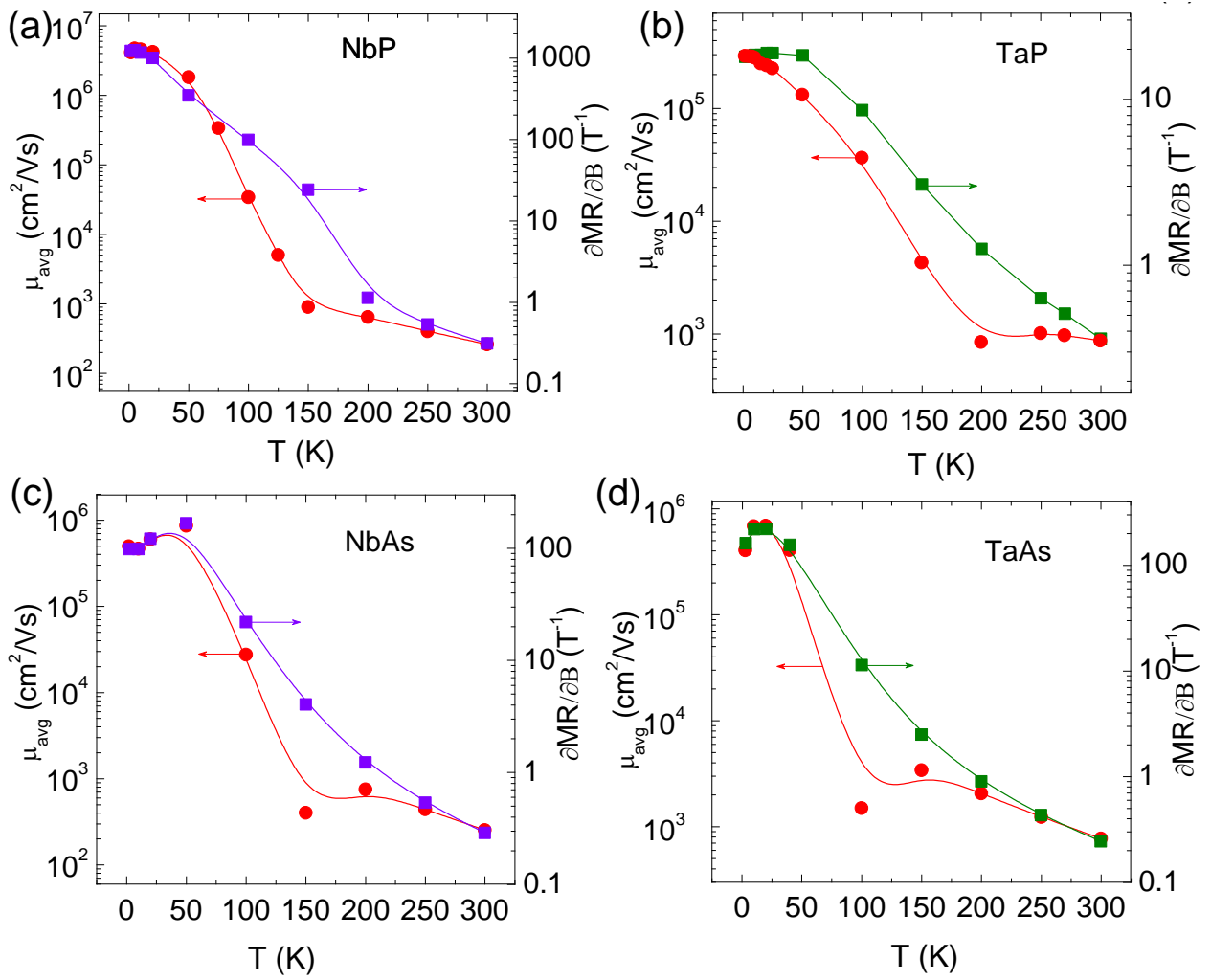

**Figure 3.** Temperature dependent average mobility (left axis) and slope of magnetoresistance (right axis) at high field in (a) NbP, (b) TaP, (c) NbAs and (d) TaAs.

Now, we focus on the origin of linear MR. The observed MR in present case is positive and linear and it has been known for a long time that regular metals exhibit a quadratic MR behavior, which saturates at high fields but the origin of linear MR is intriguing. In the nonmagnetic materials, linear MR can be realized mainly due to (i) open orbit in Fermi surface, (ii) Abrikosov theory of quantum origin [43], and (iii) Parish and Littlewood theory of classical origin [34]. The first origin of linear MR is simply ruled out because the charge carriers form closed orbits on Fermi surface. According to the quantum theory of MR, the resistivity of materials varies linearly with the magnetic field resulting in linear MR when compounds reach the quantum limit

$$\rho = \frac{1}{2\pi} \left( \frac{e^2}{\upsilon \, \varepsilon_\infty} \right)^2 \ln \varepsilon_\infty \frac{N_i}{ecn_0} B$$

where $n_0$ denotes the electron density, $N_i$ the density of scattering centers, $\varepsilon_\infty$ the dielectric constant at high frequencies and $\upsilon$ the band velocity, which is constant for a linear band in energy–momentum space. Quantum MR is characterized by non-saturating, positive values, and linearity down to low fields and, more interestingly; it is temperature independent. The MR of our semi-metal compounds also exhibit similar trends, and more importantly, it is also nearly temperature–independent up to 50 K, which indicates the presence of quantum MR. However, the condition for only one Landau level to participate in transport is $n < (eB/\hbar)^{3/2}$. If we set $n = 10^{24}$ m$^{-3}$ as obtained from the Hall measurements, we obtain B > 7 T and that indicates, the linear MR should only be present above 7 T. Interestingly, this family of compounds exhibit linear MR from very low field of 1 T, which rules out the quantum origin of linear MR in present compounds. Some typical examples of materials exhibiting quantum MR are graphene [44], Ag$_{2+\delta}$ X (X =Se, Te) [45, 46], and Bi$_2$Te$_3$ [47], with zero gap and linear Dirac dispersion. The third origin of linear MR is the classical theory which suggests that the linear MR is originated due to the mobility fluctuation. This model predicts that the MR is dominated either by the average mobility, <μ>, or by the width of the mobility distribution, <Δμ>. For highly pure compounds, the value of <Δμ> is negligible. The assumption of average mobility has been applied in a broad range of materials such as GaAs-MnAs composites [48], Heusler TIs [49, 50] and Ag$_{2+\delta}$Te [51]. To examine the role of the mobility on the linear MR, we have calculated the slope of the linear part of the MR and the mobility at different temperatures for each compound. The slope dMR/dH is plotted on right axis and the carrier mobility on the left axis as a function of temperature as shown in figure 3. The dMR/dH and the $\mu_{avg}$ show identical variation with temperature indicating that the linear MR is governed by the mobility in our compounds. A slight deviation between the temperature response of two quantities in the range $100 - 200$ K belongs to the region where the type of charge carrier switches from electron to hole. Another possibility for unsaturated MR is due to charge carriers compensation but in that case MR should fallow the parabolic ($B^2$ dependence) behavior [41, 52]. Furthermore, about the origin of MR, in principle, we do not expect any MR in compounds wherein all carriers have the same velocity, effective mass and relaxation time. But in reality, most of the compounds show MR due to different velocity, effective mass and relaxation time of carriers. According to Kohler's rule,

$$MR = \frac{\rho(B)}{\rho(0)} - 1 = \alpha \left( \frac{B}{\rho_0} \right)^{\beta}$$

where $\alpha$ and $\beta$ are the constants. The plot between MR and $\rho(B)/\rho(0)$ merge together into a single line when all carriers have single relaxation time. To investigate the role of scattering in the presence of magnetic field at different temperature, our calculated MR is plotted according to Kohler's rule as shown in figure 4 for each compound. It is clear that below 50 K, materials have same type of scattering whereas it deviates above 50 K. From the best fitting to Kohler's plot, the value of $\beta$ is found nearly to be 1.0 for all compounds that indicates the linear variation of MR with fields. In the case of parabolic MR, $\beta$ is 2.0 for charge compensated $WTe_2$ compound [52]. Kohler's plots deviate above 50 K due to the change in the majority carrier type i.e. from electron to hole.

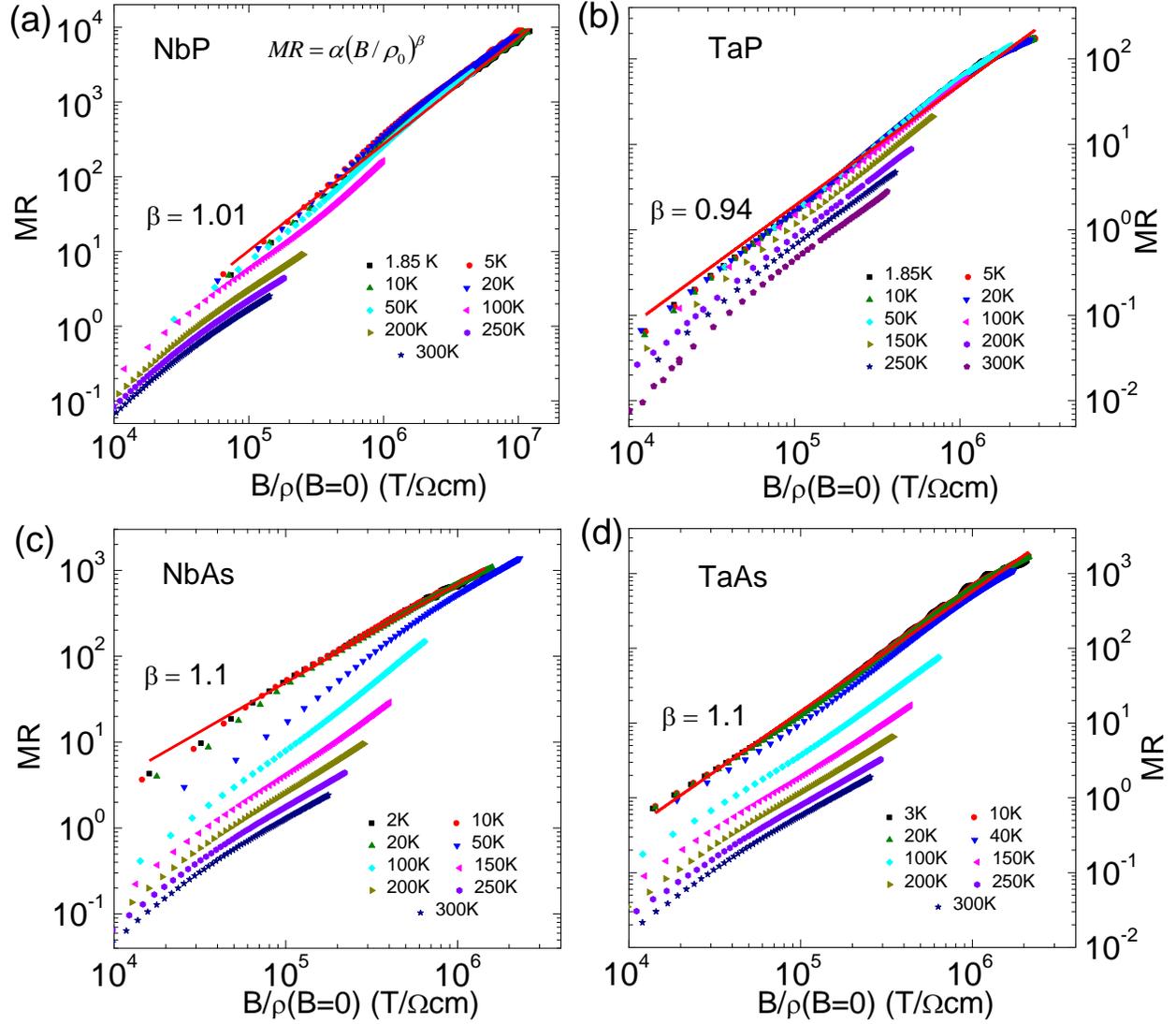

**Figure 4.** Kohler's plot between magnetoresistance (MR) and reduced field ($B/\rho$ ($B=0$)) in log scale at different temperatures in (a) NbP, (b) TaP, (c), NbAs and (d) TaAs. Value of $\beta$ is obtained from the best fitting (red line) with $MR = \alpha \left(\frac{B}{\rho_0}\right)^{\beta}$ above 50 K.

## 4. Conclusion

To conclude we show that the transition metal–monopnictides family of Weyl semimetals are very sensitive to magnetic field and show extremely high linear magnetoresistance, $8.5 \times 10^5$ % at 1.85 K for NbP and $1.5 \times 10^5$ % at 3 K in 9 T for TaAs. All the compounds, NbP, TaP, NbAs and TaAs attain very low residual resistivity reflecting high mobility of carries that plays a crucial role in the enhancement of device efficiency. Our analyses reveal that the linear MR is originated to mobility and high mobility compounds exhibit high magnetoresistance at least in this Weyl family of compounds.

## Acknowledgments

This work was financially supported by the Deutsche Forschungsgemein- schaft DFG (Project DFG-SFB 1143) and by the ERC (Advanced Grant No. 291472 Idea Heusler).